\documentclass{article}
\usepackage{wrapfig}
\usepackage{epsfig}
\usepackage{listings}
\usepackage{geometry}
\begin{document}

\title{A comparison of techniques for solving the Poisson equation in CFD}
\author{Nick Brown}
\date{}

\maketitle

\begin{abstract}
CFD is a ubiquitous technique central to much of computational simulation such as that required by aircraft design. Solving of the Poisson equation occurs frequently in CFD and there are a number of possible approaches one may leverage. The dynamical core of the MONC atmospheric model is one example of CFD which requires the solving of the Poisson equation to determine pressure terms. Traditionally this aspect of the model has been very time consuming and-so it is important to consider how we might reduce the runtime cost.

In this paper we survey the different approaches implemented in MONC to perform the pressure solve. Designed to take advantage of large scale, modern, HPC machines, we are concerned with the computation and communication behaviour of the available techniques and in this text we focus on direct FFT and indirect iterative methods. In addition to describing the implementation of these techniques we illustrate on up to 32768 processor cores of a Cray XC30 both the performance and scalability of our approaches. Raw runtime is not the only measure so we also make some comments around the stability and accuracy of solution. The result of this work are a number of techniques, optimised for large scale HPC systems, and an understanding of which is most appropriate in different situations.
\end{abstract}

\section{Introduction}
Computational Fluid Dynamics (CFD) is used in a variety of applications, from the modelling of airflow over aircraft wings to the dynamical core of atmospheric models. The fundamental component of CFD is the solving of Navier-Stokes equations and the problem can be made substantially simpler by not resolving all scales, but instead only concentrating on quantities (or scales) of interest and parameterising, or estimating, other values. Large Eddy Simulation is a computational fluid dynamics technique used to efficiently simulate and study turbulent flows.  In atmospheric science, LES are often coupled to cloud microphysics and radiative transfer schemes, to create a high resolution modelling framework that is employed to develop and test physical parametrisations and assumptions used in numerical weather and climate prediction.

In this paper we concentrate on the recently developed Met Office NERC Cloud model (MONC) \cite{easc}, an open source high resolution modelling framework that employs large eddy simulation to study the physics of turbulent flows. The goal of this model is to further develop and test physical parametrisations and assumptions used in numerical weather and climate prediction. However at its heart MONC utilises CFD for the dynamical core and the purpose of the model in this paper is as a driver for investigating the underlying computational methods. MONC replaced an existing model called the Large Eddy Model (LEM) \cite{lem} which was an instrumental tool, used by the weather and climate communities, since the 1980s for activities such as development and testing of the Met Office Unified Model (UM) boundary layer scheme \cite{lock1998}\cite{lock2000}, convection scheme \cite{petch2001}\cite{petch2006} and cloud microphysics \cite{abel2007}\cite{hill2014}. A key aspect of the LES dynamical core in MONC is the solving of pressure terms, via the Poisson equation. In the LEM model this consumed a significant portion of the overall runtime and as-such we considered it crucially important to optimise and understand the best choice of computational technique. 

The ubiquity of CFD means that lessons learnt about optimal computational approaches in one area are of relevance to other fields such as from atmospheric science in MONC to aircraft design. Section two introduces the different approaches that we developed for solving the pressure terms in MONC as well as discussing their implementation in more detail. The performance, scalability and stability of the techniques on up to 32768 processor cores of a Cray XC30 is described in section three, before some conclusions are then drawn in section four and further work discussed. 

\section{Computational approaches for solving pressure terms}
Calculation of the pressure term involves solving a Poisson-like elliptic equation \cite{lem-science} for the pressure, \emph{p}, field. This represents a pressure correction, where the dynamical core generates source terms from other aspects such as advection and buoyancy which are then used to define input to the solver. The Poisson equation is then solved which determines the current pressure at every point in the three dimensional grid, measured in Pascals. From a computational perspective not only are there are numerous ways to actually solve this equation, but many of these also contain numerous choices and configuration options. It is fair to say that this is an area of the model often less understood by the users, with them relying on some default choices, however as we will see if they tune their configuration for the actual test-case then significant improvements in the overall performance can be obtained.
\subsection{Direct FFT}
\label{section:fft}
One approach to solving the Poisson equation is based upon Fourier transformation. By performing a forward Fourier transformation into the frequency domain, we obtain a vertical ordinary differential equation that can then be solved using a tri-diagonal solver. Once completed a backwards Fourier transform reverts from the frequency domain back into to the spatial domain, this being the pressure solution. The tri-diagonal solver itself is trivial computationally and works vertically (in the \emph{Z} dimension) up and down each individual column of the system, without any interaction with neighbouring columns. This works well as long as complete columns are held on processes, which is the case (MONC decomposes in the \emph{X} and \emph{Y} dimensions only.) The computationally intensive parts of the direct FFT solver are to be found in computing the forward and backward Fourier transforms and the Fast Fourier Transform (FFT) algorithm \cite{fft} is commonly used for this. When, as in this case, working with a 3D system each process handles the dimensions separately, executing the FFT algorithm for all of the data in one dimension before moving onto the next. In this implementation we use the Fastest Fourier Transformation in the West (FFTW) \cite{fftw} library for the actual FFT computational kernel. FFTW is a fast, commonly available and open source library that implements numerous FFT algorithms. The programmer provides details of their problem in an initial planning phase, which FFTW uses to select the most appropriate kernel to execute. The idea being that the planning and selection of kernel(s) is expensive, but only done once (such as during an initialisation step) as the plan can then be reused time and time again (for instance every timestep.)

When computing Fourier transforms in parallel systems typically one has a choice between splitting the system up (decomposing the domain) in one or two dimensions. Decomposing in one dimension, known as slab, is  simpler to implement but does limits the amount of parallelism available in contrast to a two dimensional, or pencil, decomposition. FFTW supports parallelism through a one dimension slab decomposition, however this is notoriously inefficient. Therefore in our code we have leveraged FFTW for the actual computational kernels but implemented our own, pencil decomposition and parallelism to distribute the system amongst the processes in two dimensions. Hence FFTW is responsible for the computation and our own code the decomposition and communication.

\begin{wrapfigure}{l}{0.35\textwidth}
	\centering
	\includegraphics[width=0.25\textwidth]{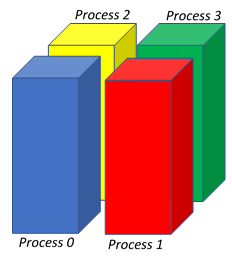}
	\caption{Vertical (Z-based) column orientation of data}
	\label{fft:one}
\end{wrapfigure}

\begin{wrapfigure}{r}{0.35\textwidth}
	\centering
	\includegraphics[width=0.25\textwidth]{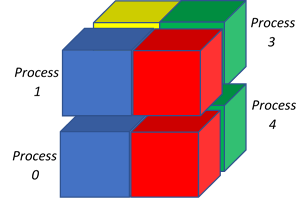}
	\caption{Y dimension orientation of decomposition}
	\label{fft:two}
\end{wrapfigure}

The pressure field, \emph{p}, that we are working on is initially distributed as in figure \ref{fft:one}. The domain is decomposed between processes in two dimensions, with columns resident on a process. In this example there are four processes, each in possession of a unique set of columns. Communication is then issued to transpose in the \emph{Y} dimension, where each process communicates with a subset of processes (all processes in the dimension's specific direction) to gather together data and distribute its own to form another pencil in this dimension. This can be thought of as a parallel transposition in that dimension, and the result of this communication is illustrated in figure \ref{fft:two}. After this transposition, data is now allocated to processes in rows rather than columns. For each row residing on a process the forward Fourier transform is performed by calling into FFTW which executes the appropriate internal kernel as per its plan.

\begin{wrapfigure}{l}{0.35\textwidth}
	\centering
	\includegraphics[width=0.25\textwidth]{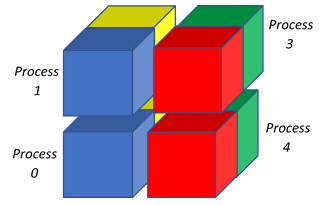}
	\caption{X dimension orientation of decomposition}
	\label{fft:three}
\end{wrapfigure}

Once work in the \emph{Y} dimension has completed then a further transposition and FFT in the \emph{X} dimension must be performed. As before, communication is required with processes in this dimension, in order to gather together blocks forming a pencil in \emph{X} as illustrated by figure \ref{fft:three}. Once again a forward FFT is performed for all data in this dimension. 

At this point we have completed the forward Fourier transform for the 3D data and the pressure field, \emph{p}, is in the frequency domain. Effectively we have an ODE which our simple tri-diagonal solver can operate upon. However, a significant restriction of our solver is that it must work in vertical columns but the data, as per figure \ref{fft:three}, is split up and distributed in the \emph{X} dimension. As the algorithm requires data to be oriented in the vertical, \emph{Z}, dimension further communication is required to gather the data back to its original orientation and we do this by reverse transposing, from the \emph{X} dimension (figure \ref{fft:three}) to the \emph{Y} dimension (figure \ref{fft:two}) and then to the \emph{Z} dimension (figure \ref{fft:one}).

Once the tri-diagonal solver has executed we need to transform from the frequency domain back into the time domain. This requires a backwards 3D Fourier transform and first we must transpose back to the \emph{X} dimension via the \emph{Y} dimension. A backwards FFT is executed by FFTW in this \emph{X} dimension before transposing to the \emph{Y} dimension and performing another backwards FFT in \emph{Y}. Lastly a transposition from \emph{Y} to \emph{Z} is executed which gives us the pressure field in the time domain and holds the solution to the pressure equation.

From our description here it can be seen that there is significant communication between the processes when performing all these data transpositions. Not only this, but because we are splitting the data into blocks that are then distributed between numerous processes, this can result in many small messages between many processes which is known to be inefficient. More general optimisation techniques, such as overlapping communication with computation could offset this inefficiency but these are not applicable here because we must be in possession of all a dimension's data before performing the Fourier transform. In short, due to the cost of communication, it is widely accepted that FFTs are not scalable to large numbers of processes.

An uneven decomposition of data is where the global domain does not divide evenly between processes and so they hold different numbers of elements. This can add considerable complexity to the code but is an important case for us to handle. We found that the best way to implement this has been to follow a similar initial planning approach that FFTW exposes. Instead of selecting computational kernels our code is determining a \emph{plan} for decomposition and communication. Specifically this plan is made up of a number of transpositions, each transposition containing specifics of what local data needs to be sent and what remote data to receive and then where to place it. In the plan, associated with the set of transpositions, are a number of buffers. These are needed for communication and we aim to maximise re-use of these buffers for memory efficiency reasons. Whilst the resulting implementation is inevitably complex, the fact that performing a forward Fourier transform and then a backwards transform will return the original values means that testing correctness is trivial. By temporarily disabling the tri-diagonal solver we have been easily able to test that the input values match output values from the transpositions to provide confidence that the decomposition and communication is correct irrespective of the system size and number of processes.

Previously the model utilised an old FFT implementation which, written in Fortran 66, only implemented slab decomposition and exhibited very poor performance due to naive parallel communications and a poor FFT kernel. By reimplementing and leveraging FFTW for the kernel with our own pencil decomposition and communication routines we were able to obtain a very significant improvement in overall performance in MONC compared to the LEM.

\subsection{Iterative solver}
The second way of solving the Poisson equation is by an iterative method where the system starts off with an initial guess and makes process towards the final answer via a series of iterations. A major advantage this technique holds for MONC is that it does not require the lateral periodic boundary conditions imposed by the direct FFT solver and so extends the flexibility of the model in general. We are looking to solve an equation of the form $Ax=b$, where \emph{A} is the matrix, \emph{x} is the target solution and \emph{b} is the right hand side (RHS). Mapping this to the Poisson equation, that we are solving here for pressure terms, one obtains $\bigtriangledown^2x=b$. Whilst there are many variants of iterative methods, such as Jacobi and Gauss-Seidel, Krylov subspace methods form the basis of the most efficient iterative solvers. There are numerous Krylov based solvers such as Conjugate Gradient (CG), BIConjugate Gradient STABbilized (BiCGStab) and Generalized Minimal RESidual (GMRES.) Each of these varies in its performance and stability for specific problems, so the trick is to pick the most appropriate solver for the equation in question. Before starting to solve the system it is common to pre-condition the matrix in order to lower its condition number. Broadly speaking the condition number can be used to determine how complex a system will be to solve, and a high condition number will result in more iterations of the solver. Therefore running an algorithm on the matrix to lower the condition number, and hence reduce the number of solver iterations by making the system easier to solve, can be of significant performance benefit.

\begin{wrapfigure}{l}{0.35\textwidth}
	\centering
	\includegraphics[width=0.25\textwidth]{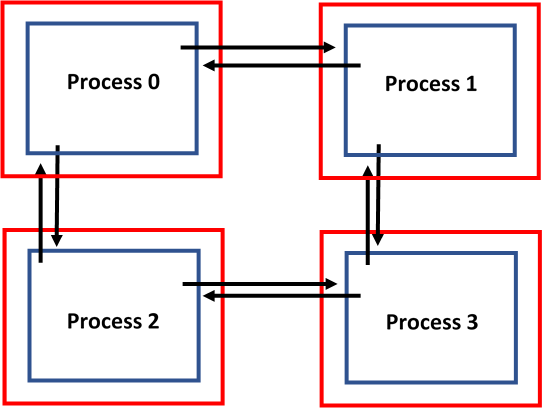}
	\caption{Iterative solver halo-swapping}
	\label{fig:hs}
\end{wrapfigure}

Unlike the direct FFT approach, iterative methods proceed in iterations. Instead of resulting in the exact solution they will terminate once a specific accuracy has been met and the relative residual, i.e. by how much the error of the solution has decreased, is a common measure and used in this work. For the calculation of each grid point in the system one requires the neighbouring points. This is simple in the middle of the local domain, as it just requires a local memory read, but on the boundaries communication is required with the neighbouring process to obtain the needed values. For efficiency a technique, \emph{halo swapping}, is used which is illustrated in figure \ref{fig:hs} where a halo of width one element surrounds the local domain on each process. After an iteration, values are exchanged with all neighbours (in this case a maximum of four neighbours because we decompose in two dimensions) en-mass. This is optimal because fewer, larger messages are far more efficient than many smaller ones, and it also provides the possibility of overlapping some computation with communication. In contrast to the direct FFT approach, halo swapping communication required by the iterative solver is more regular, localised and easier to optimise. In short the communication pattern of the iterative solver should provide significant performance benefits over communication of the FFT solver when considering larger core counts.

\subsubsection{PETSc iterative solver}
The Portable Extensible Toolkit for Scientific computation (PETSc) \cite{petsc} is a general purpose framework for solving linear systems. Combining a variety of solvers, pre-conditioner and methods for parallelisation, one can define their matrix and RHS using the provided data structures and abstractions. This is handed to PETSc which then executes the actual solve and provides the solution to the user. For parallelisation a Distributed Memory (DM) interface is provided. This enables the programmer to define general details about their problem and desired parallelism, with the library then making the lower level choices and implementing the physical mechanism such as communication. A major advantage of PETSc is the availability of a rich collection of solvers and pre-conditioners. As such it is trivial to experiment with different solution choices simply by setting general options in the code to enable experimentation to find the most appropriate solver for performance and stability.

Whilst it is possible to run PETSc in a matrix-less manner, i.e. without physically constructing the matrix or RHS, this requires significantly more work and neither the existing solvers nor pre-conditioners can be used directly with this. Therefore in MONC we go with the more common approach of explicitly constructing the matrix and RHS in memory. Listing \ref{lst:petscsetrhs} illustrates the Fortran subroutine called by PETSc to setup  the RHS which is based upon source terms which have already been computed for the current timestep. The distributed matrix is first retrieved and then the current processes' global start grid point \emph{xs}, \emph{ys} and \emph{zs} along with the local data size in each dimension \emph{xm}, \emph{ym} and \emph{zm} are determined. A Fortran pointer to the underlying RHS array is retrieved and then the \emph{copy\_data\_to\_petsc\_pointer} subroutine, which we developed, will copy the pressure source terms \emph{p\_source} (a global variable) into this array. 
\begin{lstlisting}[frame=lines,caption={PETSc RHS for each solve},label={lst:petscsetrhs}]
subroutine compute_RHS(ksp, b, dummy, ierr)
PetscErrorCode :: ierr
KSP :: ksp
Vec :: b
integer :: dummy(*)

DM dm
PetscScalar, pointer :: xx(:)
PetscInt :: zs, ys, xs, zm, ym, xm

call KSPGetDM(ksp, dm, ierr)
call DMDAGetCorners(dm, zs, ys, xs, zm, ym, xm, ierr)
call VecGetArrayF90(b, xx, ierr)
call copy_data_to_petsc_pointer(xx, zs, ys, xs, zm, ym, xm, p_source)
call VecRestoreArrayF90(b, xx, ierr)
end subroutine compute_RHS
\end{lstlisting}

Matrix generation is also done via a procedure callback, omitted for brevity, which is called on the first timestep only as once the matrix is initialised it remains unchanged. This matrix implements the Poisson equation in a standard manner and has been tested for correctness in comparison to the FFT direct solver. Listing \ref{lst:petsccall} illustrates the calling of PETSc at each timestep to perform the pressure solve. The \emph{p\_source} global variable is set to be the current source terms which are held in the \emph{p} (pressure) data array and this is then read in the \emph{compute\_RHS} procedure called from PETSc. Execution of the \emph{KSPSolve} procedure actually performs the solve and, once this has completed, the \emph{KSPGetSolution} procedure retrieves the solution as a PETSc vector which is then read and copied back into the \emph{p} array. The \emph{prev\_p} global variable is set to this value which will be initial conditions for the pressure solve at the next timestep.

\begin{lstlisting}[frame=lines,caption={Executing PETSc iterative solver},label={lst:petsccall}]
p_source=current_state%p%data(z_start:z_end, y_start:y_end, x_start:x_end)
call KSPSolve(ksp, PETSC_NULL_OBJECT, PETSC_NULL_OBJECT, ierr)
call KSPGetSolution(ksp, x, ierr)
call VecGetArrayReadF90(x, xx, ierr)

call copy_petsc_pointer_to_data(xx, z_start, z_end, y_start, y_end, x_start, x_end, current_state%p%data)
call VecRestoreArrayReadF90(x, xx, ierr)
prev_p=current_state%p%data(z_start:z_end, y_start:y_end, x_start:x_end)
\end{lstlisting}

In contrast to the direct FFT solver, leveraging PETSc for an iterative approach is considerably simpler from a code perspective. This is mainly due to the mature support of the library, but it also should be noted that because the neighbour to neighbour communication pattern of iterative solvers is more regular and far simpler that that required by the Fourier transform, it is easier to abstract in a common library.

\subsubsection{Bespoke solver implementation}
The benefit of the PETSc based approach is that numerous existing solvers and preconditions are made immediately available, however explicit bespoke implementations of the CG and BiCGStab methods were also developed for solving the Poisson equation in MONC. A benefit to this bespoke approach is that we can tune our implementations to explicitly match the problem, rather than having to rely on PETSc's general implementation. Our approach is matrix-less, i.e. the matrix and RHS are not explicitly constructed and as-such this utilises less memory than the PETSc implementation. This implementation post-conditions rather than pre-conditions where we effectively solve $AM(Nx)=b$, where \emph{N} is the inverse of \emph{M}. This post-conditioner has been designed specifically for our problem and the use of a post-conditioner means that there is flexibility to change \emph{M} at every iteration in theory which might improve performance. When relying on pre-conditioning this isn't the case since the RHS has been altered and so the LHS needs to stay consistent.

The specific calculation procedures of the solvers, such as calculating Ax (the matrix vector product), is hard coded to the problem by physically implementing calculations necessary to the Poisson equation in code. It would therefore be impossible to target these solvers at other equations without a significant re-implementation of the code but this does mean that we can heavily tune for the problem in hand. One example of this is in the avoidance of communication, where we can combine multiple inner products, each requiring a collective reduction, into one. We are also able to overlap communication with compute by initiating halo swapping in a non-blocking manner. Whilst this is in progress we then start working on the inner local domain calculations which do not require neighbouring data. Only once all the inner value calculations have completed, which represents the vast majority of the domain, will the code then wait for halo swapping communications to complete and handle boundary values. The intention is that whilst the relatively long running inner computations have been in progress the halo-swapping has completed and-so the cost of communication is greatly reduced because the process can continue doing useful work instead of stalling whilst it completes.

\section{Comparison}
As mentioned in section \ref{section:fft}, traditionally a very naive FFT implementation had been used to solve the Poisson equation in the LEM. This exhibited very poor performance and would not scale beyond a couple of hundred processes. Our initial plan was to address this by utilising an iterative solver. However at the same time we also re-implemented the FFT solver as discussed in section \ref{section:fft} which in itself resulted in very significant improvements to the runtime. Even though it is significantly improved, the FFT solver still suffers from two distinct disadvantages; firstly the significant, complex and inefficient communication required, and secondly the fact that it is a direct method and hence provides the exact solution every time. Therefore it is useful to provide an alternative to this FFT approach and to understand the different performance and scalability characteristics that we might expect from them. It is very possible that different configurations of the model will favour different approaches and/or methods, therefore providing a choice is positive.

All performance tests in this section have been run on ARCHER, a Cray XC30 which is the UK national supercomputer. In all cases the Cray compiler was used and the results presented are the average of three runs. A standard dry boundary layer MONC test case has been used to drive the solvers which is concerned with the atmosphere at the interface between two different air masses.

\begin{figure}		
	\begin{center}
		\includegraphics[scale=0.64]{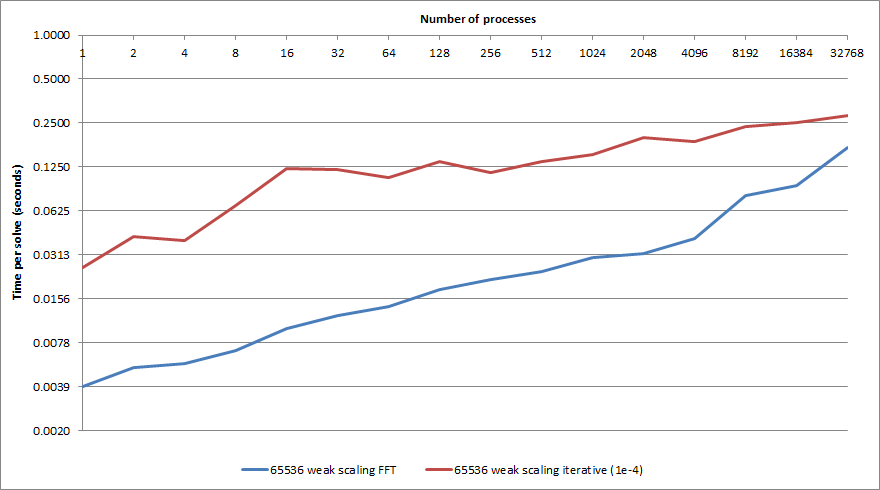}
	\end{center}
	\caption{Performance of FFT and iterative solver as the number of processes is scaled}
	\label{fig:fft-iterative}
\end{figure}

Figure \ref{fig:fft-iterative} illustrates the performance of the FFT and PETSc iterative solver. In this experiment we are weak scaling with 65536 local grid points and hence 2.1 billion global grid points at 32768 cores. The PETSc solver is leveraging the BiCGStab solver with ILU preconditioner. It can be seen that the FFT solver is significantly faster than the PETSc iterative solver at smaller core counts, but as one scales up the differences diminish. Hence it can be seen that the direct FFT solver involves less computation but more communication. Results of figure \ref{fig:fft-iterative} are averaged over the entire run which involves significant model spin-up time where the field is varying considerably between timesteps. This variation causes the number of iterations of the iterative solver to be high whereas later on in the run, when the model settles down, then the number of iterations becomes minimal because a steady state has been reached. Conversely the FFT solver requires the same amount of computation regardless, this is advantageous where there is significant variance but less so for stable systems. 

A major benefit when utilising an iterative solver in contrast to a direct solver is that we can set the termination residual and therefore explicitly choose how accurately to solve the system. Table \ref{tbl:iterativeperformance} illustrates executions of the iterative solver, solving to different relative residual accuracies, against the direct FFT solver. The \emph{Mean difference} column illustrates the average difference per field element the pressure values calculated by the iterative solver and that by the direct solver. The focus here is on computation rather than communication, hence the timing results are over executions of the test case on one core only and the FFT solver took on average 3.78e-3 seconds per timestep. It can be seen that solving iteratively to 1e-2 is significantly faster than a direct FFT solve, although there is the most significant difference in terms of solution accuracy. It is up to the user of MONC to determine the exact termination residual to solve to for their own configurations and our general suggestion is that somewhere between 1e-2 and 1e-4 is most appropriate. The fact that one can leverage the target residual accuracy to considerably reduce the overall solver time is important for users to understand and for some, especially large coarse grained, systems a high level of accuracy in the pressure calculation is not required.

\begin{table}[h]
	\centering
	\begin{tabular}{ | c | c | c | }
		\hline
		Relative termination residual \quad & \quad Mean difference \quad & \quad Average time per solve (s) \\
		\hline
		1e-2 \quad&\quad 1.68e-5 \quad&\quad 1.06e-3 \\
		1e-4 \quad&\quad 9.08e-7 \quad&\quad 5.30e-3\\	
		1e-6 \quad&\quad 4.52e-9 \quad&\quad 2.71e-2\\
		1e-8 \quad&\quad 9.91e-11 \quad&\quad 6.27e-2\\
		FFT  \quad&\quad N/A \quad&\quad 3.78e-3 \\
		\hline
	\end{tabular}
	\caption{Iterative solver accuracy in comparison to FFT solver and average time per solve}
	\label{tbl:iterativeperformance}
\end{table}

Table \ref{tbl:solvers} illustrates the performance difference between different solvers in both PETSc and our bespoke implementation. The ILU pre-conditioner has been used for all PETSc runs as this was found to be the optimal choice for our problem. It can be seen that GMRES, which is PETSc's default solver, is marginally slower than  BiCGStab in PETSc. Our own implementation of BiCGStab is faster than PETSc and the fastest is our bespoke implementation of CG. Whilst the number of solver iterations varies dramatically during the model run, PETSc's GMRES tends to involve the greatest number and our bespoke CG implementation the fewest iterations. The number of iterations between PETSc's BiCGStab and our bespoke BiCGStab is roughly similar (even though a different approach to lowering the condition number of the matrix is adopted.) In all it is felt that our bespoke solvers out performs PETSc due to the optimisations we have been able to apply due to the more specialist implementations. This is both in terms of the targetted post conditioning and also communication optimisation (avoidance and overlapping.)

In terms of solver stability the default GMRES solver of PETSc is able to solve the pressure equations of all our test-case configurations. PETSc's BiCGStab solver works for most test-cases, but can become unstable with some non-symmetric systems. By contrast our own bespoke BiCGStab solver is stable for all systems and the CG solver requires symmetric systems which represents the smallest subset of the test-case configurations. It is important for the user to understand whether CG would be an appropriate choice for their problem because, if so, then this will result in the lowest runtime.

\begin{table}[h]
	\centering
	\begin{tabular}{ | c | c | c | }
		\hline
		Solver \quad & \quad Average time per solve (s) \\
		\hline
		GMRES (PETSc) \quad&\quad 5.68e-3 \\
		BiCGStab (PETSc) \quad&\quad 5.30e-3\\	
		BiCGStab (bespoke) \quad&\quad 4.98e-3\\
		CG (bespoke) \quad&\quad 4.58e-3\\
		\hline
	\end{tabular}
	\caption{Performance of iterative solvers}
	\label{tbl:solvers}
\end{table}

In addition to the runtime there is also the question of memory usage. The runs in this paper have focussed on weak scaling with 65536 local grid points per process. In reality this represents a rather small local domain and can easily fit into RAM (2GB per core on ARCHER.) Whilst this is generally a recommended local domain size, some users do want to run with significantly larger local domain sizes. Therefore it is important to consider the memory requirements of these different techniques as this will ultimately limit the amount of local data that can be stored. Our PETSc iterative solver requires substantially more memory than the FFT or bespoke solvers, this is not only due to the fact that we explicitly construct the matrix and RHS in memory but also because PETSc itself abstracts parallelism and operations through matricies, vectors and index sets. Hence internal to PETSc a number of these data structures need to be constructed which results in additional memory overhead. The bespoke solver implementation is the most memory efficient, only requiring a small number of buffers for the halo swapping communications, quite closely followed by the direct FFT solver which requires slightly more memory. In reality a user will see very little difference in memory limitation between the FFT and bespoke iterative solvers, but for larger local system sizes they will be impacted by the memory requirements of the PETSc approach.

\section{Conclusions}
In this paper we have considered two different computational techniques for solving the Poisson equation as part of the pressure correction step of the MONC model's dynamical core. It has been seen that there is a trade-off between computation and communication. The cost of computation associated with a direct FFT method is lower, but the cost of communication more significant. At lower core counts our direct FFT approach is significantly faster because communication is not a significant factor, but as we scale up to greater numbers of cores then the difference diminishes. In terms of runtime performance there is a slightly increased cost of using the general PETSc framework instead of the bespoke solvers, however implementation of the former was far quicker and the benefit of PETSc is that we can trivially switch between solvers and pre-conditioners. Therefore, whilst currently a bespoke iterative solver implementation targetted at the specific problem is advantageous, as the large PETSc community improve the library and develops further methods then these might very well out perform our own solvers. 

In truth we were surprised that the direct FFT solver scaled as well as it did. However one should bear in mind that Cray machines, with their Aries routing technology, have a very efficient interconnect between the nodes hence diminishing the impact of communication inefficiencies. It is likely that scalability on HPC machines inter-connected with other technologies, such as Infiniband, would exhibit poorer scaling. One should also bear in mind that as we progress towards exa-scale and look to leverage these methods to solve challenging scientific and engineering problems, 32768 processes will not be considered a large amount of parallelism. From figure \ref{fig:fft-iterative} it is fairly clear that if one was to target a hundred thousand processes and beyond, which are future plans for MONC, then the iterative approaches would be preferable irrespective.

In terms of further work there is the possibility of optimising the communication in the direct FFT solver by eliminating the need to \emph{go back} to a column based (Z) decomposition for the tri-diagonal solver. If this could be eliminated then it would half the overall amount of communication which would be a significant improvement. Whilst modifying the tri-diagonal solver away from a column based approach might be difficult, in our current code instead of transposing directly from an X decomposition to a Z decomposition, the code goes via the Y dimension requiring extra communication. This was done for simplicity and to reuse the transposition implementation, but removing this intermediate step would reduce the overall amount of communication, albeit at the cost of code complexity.

\end{document}